\documentstyle[fleqn,12pt]{article}

\renewcommand{\baselinestretch}{1}
\renewcommand{\arraystretch}{1}
\small\normalsize

\newcommand{\be}{\begin{equation}}
\newcommand{\ee}{\end{equation}}

\newcommand{\eps}{\varepsilon}
\newcommand{\om}{\omega}
\newcommand{\bz}{_{\scriptstyle\rm BZ}}
\newcommand{\mtrx}[1]{\underline{\underline{#1}}}

\begin{document}
\vspace{-45mm}
\begin{flushright}
K\"oln, ITP/94, cond-mat/9408061 \\
\today
\end{flushright}

\vspace{-6mm}

\begin{flushleft}
\renewcommand{\baselinestretch}{2} \small \normalsize
{\LARGE Conductivity of interacting spinless fermion systems
 via the high dimensional approach}
\end{flushleft}

{\vspace{-4mm}}

\begin{flushleft}
 {\Large G\"otz S. UHRIG}
\end{flushleft}

\begin{flushleft}
\normalsize \em  Inst.\ f.\ Theor.\ Physik,
 Universit\"at zu K\"oln, 50937 K\"oln, Germany
\end{flushleft}

\vspace{-4mm}

\begin{abstract}
\renewcommand{\baselinestretch}{1}
\renewcommand{\arraystretch}{1}
\small\normalsize
 Spinless fermions with repulsion are treated non-perturbatively by classifying
the
diagrams of the generating functional $\Phi$ in powers of the
inverse lattice dimension $1/d$.
The equations derived from the first two orders are evaluated
on the one- and on the two-particle level.
The order parameter of the AB-charge density wave (AB-CDW) occurring
at larger interaction is calculated in $d=3$.
The Bethe-Salpeter equation is evaluated for the
conductivity $\sigma(\om)$ which is found to have two peaks within the
energy gap $2\Delta$ in the AB-CDW:
 a remnant of the Drude peak and an excitonic
resonance. Unexpectedly, $\sigma_{\rm\scriptscriptstyle DC}$ does not vanish
for $T\to 0$
\end{abstract}

\vspace{5mm}
%noindent
%Keywords: conductivity, charge-density wave, $1/d$ expansion, Bethe-Salpeter
%%equation

\newpage
\renewcommand{\baselinestretch}{1}
\renewcommand{\arraystretch}{1}
\small\normalsize
The limit of infinite dimensions $d\to \infty$
for itinerant fermion systems \cite{metzn89,mulle89}
 has proven very useful for the
understanding of strongly interacting electron systems
\cite{vollh93}. The research
focusses on one-particle properties in the strict
limit $d=\infty$. There exist also some results on
transport properties \cite{khura90,prusc}.
 The present contribution
extends the knowledge on the AC- and DC-conductivity including
$1/d$-corrections in the model of spinless fermions with
 repulsive interaction.
The results cover and the non-symmetry broken (homogeneous) phase
and the AB-CDW occurring on
 bipartite lattices.
The techniques used in the description of the latter
are certainly transferable to
 other symmetry broken phases in related models.

The Hamiltonian  of the
 spinless fermion model at half-filling ($n=0.5$)
 is given in second quantisation
\begin{equation}
\hat H = -\frac{t}{\sqrt{2d}}\sum\limits_{<i,j>}
\hat c_i^+ \hat c_j ^{\phantom{+}}
+ \frac{U}{4d}\sum\limits_{<i,j>} \hat n_i \hat n_j -\frac{U}{2} \sum
\limits_{i} \hat n_i   \ .
\end{equation}
Scaling with the inverse dimension $1/d$ of the
considered hypercubic lattice
 is performed to ensure the continuity
 of the limit $d\to \infty$.  The model is appropriate for the description of
strongly
polarised fermions where the band of one of the spin species is
completely filled (for references on the model see ref.\ \cite{halvo94}).
 In the strict limit $d=\infty$ the perturbation series
in the interaction can be resummed since only the Hartree terms contribute
\cite{mulle89}. This transparent situation provides a good starting point
for the calculation of $1/d$-corrections \cite{halvo94,mydiss}.
In view of the
preservation of conservation laws, it is, however,
not trivial to include $1/d$-corrections. At least for the considered case
it was shown in detail that the formalism of Baym/Kadanoff
 \cite{baym} avoids inconsistencies \cite{mydiss}. In this formalism
 the self-energy $\Sigma(1;2)$
($1$ being shorthand  for $(i_1,\tau_1)$ etc.)
and the kernel of the Bethe-Salpeter equation
$\Xi(1,2;3,4)$ are derived from a generating functional $\Phi$ according to
$\Sigma(1;2)=\delta\Phi/\delta G(2;1)$ and $\Xi(1,2;3,4)=
\delta\Sigma(1;3)/\delta G(2;4)$. The exact
$\Phi$-functional is the sum of all closed skeleton diagrams.

To obtain an expansion in the inverse dimension $1/d$
the diagrams belonging to
$\Phi$ are classified in powers of $1/d$. Truncation
after the desired order yields an approximate $\Phi_{\rm N}$
and hence an approximation for $\Sigma$ and $\Xi$. In the present work
the two leading orders ${\cal O}(1)$ and  ${\cal O}(1/d)$ are kept. The
resulting $\Phi_{\rm N}$ is shown in Fig.\ 1.
%Figure 1 should be place here
 Halvorsen, Uhrig and Czycholl
set up and evaluated the ensueing system of self-consistent equations
 \cite{halvo94}.
 Here it is sufficient to know that the self-energy consists of two
site-diagonal
terms $\Sigma^{\rm H}_i$ (Hartree) and $\Sigma^{\rm C}_i$ (local correlation)
and the Fock-term $\Sigma^{\rm F}$ which links adjacent sites
 and renormalises
thereby the hopping $t \to t't := t-\Sigma^{\rm F}\sqrt{2d}$.

In the homogeneous
phase the local self-energy is constant. In the AB-CDW the local self-energy
alternates from site to site. The order parameter $b$ of this phase is the
particle density difference between a specific site and the average density
$b:= |\langle \hat n_i\rangle -n|$, i.e.\ the density is above (below) average
on the
sites on sublattice A (B) of the bipartite hypercubic lattice.
In Fig.\ 2 the generic behaviour of $b(T)$ is shown and
%Figure 2 should be place here
compared to the results of a Hartree- and a
Hartree-Fock calculation. The inclusion of $1/d$ terms does not change the
qualitative form
of the curve. The quantitative changes are also quite moderate.
 From this and from a quantitative comparison
with the exact results in $d=1$ \cite{halvo94}
one can deduce that the $1/d$ expansion yields a good
approximation in $d=3$.

 At $n=0.5$ the spontaneous
symmetry breaking implies an energy gap $2\Delta$ at the Fermi level.
This gap is lowered by the $1/d$ terms
 as is the order parameter $b$ in Fig.\
2. At $T=0$
the imaginary part of the local self-energy displays a gap which is
exactly three times $2\Delta$ \cite{halvo94}.
 This stems from the fact that due to the
gap in the one-particle spectrum
 a minimum energy  of $6\Delta$ is required for an inelastic scattering
process. This phenomenon implies the existence of undamped quasi-particles
for energies in the interval $(\Delta,3\Delta)$ measured from the
Fermi level \cite{halvo94,mydiss}. They lead to
interesting effects in the
conductivity (see below).

To calculate two-particle quantities the Bethe-Salpeter equation with
the kernel $\Xi$ must be solved which is
 a very complicated task. For the
conductivity $\sigma(\om) =\sigma_1(\om)+\sigma_2(\om)$,
 the current-current response function
$\chi^{\sf JJ}(\om)$ must be known since $\sigma_1=-i\langle\hat T\rangle/(\om
d)$ and
 $\sigma_2=-\chi^{\sf JJ}/\om$ holds
($\langle\hat T\rangle$ is the kinetic energy). Due to the odd parity of the
current vertex {\sf J} in $\bf k$-space
 only a part of the Bethe-Salpeter equation contributes
to $\chi^{\sf JJ}$ \cite{khura90}. This leads in the present case to a
tractable geometric series for $\chi^{\sf JJ}$ which is
shown in Fig.\ 3. The vertical interaction lines
result from the Fock diagram which is second in Fig.\ 2. They represent
the first vertex corrections in an expansion in $1/d$.
%Figure 3 should be placed here
The series in Fig.\ 3 leads to $\chi^{\sf JJ}=\chi^{\sf JJ}_0/(1+U\chi^{\sf
JJ}_0/2)$
if $\chi^{\sf JJ}_0$ is defined as the simple bubble of dressed propagators
(first
diagram in Fig.\ 3). Explicitly, $\chi^{\sf JJ}_0$ is given in the
homogeneous phase by
\be
\label{con1}
\chi^{\sf JJ}_0(i\zeta_m) =
\frac{2T}{d} \sum\limits_{\zeta_\nu-\zeta_\lambda=\zeta_m} \;\int\limits\bz
\frac{\sin^2(k_1)}
{\left(w_{\nu}-t'\eps({\bf k})\right)
\left(w_\lambda-t'\eps({\bf k})\right)}
\frac{dk^d}{(2\pi)^d} \ ,
\ee
where  $w_{\nu/\lambda}:=i\zeta_{\nu/\lambda}-\Sigma(i\zeta_{\nu/\lambda})$
with
roman indices for bosonic Matsubara-frequencies and greek indices for
fermionic.
The dispersion $\eps({\bf k})$ is $t\sqrt{2/d}\sum_{i=1}^d
\cos(k_i)$.
The sum over the ${\bf k}$-vectors in the Brillouin-zone BZ can be carried out
with
a density $\rho_{\rm L}^0(\om):=\int\bz (\sin k_1)^2 \delta(\om-\eps({\bf k}))
dk^d/(2\pi)^d$. If $\rho^0(\om)$ is the usual DOS one may find $\rho_{\rm
L}^0(\om)$
from the relation $\rho^{0}(\om)
=-(2/\om)\partial{\rho^{0}_{{\rm L}}}/\partial{\om}$
\cite{mydiss,uhrig94c}. In the AB-CDW, $\chi^{\sf JJ}$
is the $(1,1)$-coefficient of a $2\times2$
matrix  $-2\left(1+U\mtrx{A}\right)^{-1}/U$
since already the one-particle propagators are $2\times 2$ matrices.
 The quasi-particles
 at wave vector ${\bf k}$ couple with those at ${\bf k}+{\bf Q}$
where ${\bf Q}=(\pi,\pi,\ldots,\pi)^\dagger$. The matrix
$\mtrx{A}=[[A_1,A_3],[A_3,A_2]]$ is the piece of diagram
between two wavy lines in Fig.\ 3; its coefficients are given by
\be\label{con2}
A_i =
\frac{T}{d}\sum\limits_{\zeta_\nu-\zeta_\lambda=\zeta_m}
\; \int\limits_{-\infty}^\infty \frac{I_i\; \rho^{0}_{\rm L}(\eps)\;d\eps}
{\left(w_\nu^2- (t'\eps)^2-\Delta^2_\nu\right)
\left(w_\lambda^2- (t'\eps)^2-\Delta^2_\lambda\right)}
\ , \ee
where
$\Delta_{\nu/\lambda}=
(\Sigma_{\rm A}(i\zeta_{\nu/\lambda})-\Sigma_{\rm B}(i\zeta_{\nu/\lambda}))/2$
 is half the difference of the local self-energy on the A- and the
B-sublattice.
The average self-energy is still denoted by $\Sigma$.
The numerators depend on $i$: $I_1=w_\nu w_\lambda+(t'\eps)^2
-\Delta_\nu\Delta_\lambda$; $I_2=I_1-2(t'\eps)^2$ and
$I_3=w_\nu \Delta_\lambda
-w_\lambda\Delta_\nu$ \cite{mydiss,uhrig94c}.

The evaluation of (\ref{con1}) and (\ref{con2}) using the exact
3-dimensional densities $\rho^0$ and $\rho^0_{\rm L}$ leads to
Fig.\ 4.
%Figure 4 should be placed here.
In the homogeneous phase at higher $T$ (curve a) the main feature of
the real part of
$\sigma_{\rm\scriptscriptstyle AC}$ is the Lorentzian Drude peak at $\om=0$.
Its width is proportional to $-{\rm Im}\  \Sigma(\om=0)$, i.e.\  to $T^2$.
In the AB-CDW at lower $T$ (curve b),
 the energy gap is clearly visible.
Yet there remains a peak of finite height at $\om=0$, of which the width, and
hence the weight, tend exponentially to zero for $T\to 0$.
 The edge, where
dissipation sets in ($\om\approx 1.15$), is located at $2\Delta$.
It is preceded by a pronounced Lorentzian
 which becomes a $\delta$-peak
of finite weight at $T=0$ (curve c). Mathematically, this peak results from
the divergence of the geometric series in Fig.\ 3. Physically, the
interaction lines in Fig.\ 3 represent a continued attraction between a
quasi-particle and a quasi-hole forming an exciton. Due to the
binding energy of the latter the corresponding resonance
 can be excited at energies a little lower than $2\Delta$.

%Figure 5 should be place here
In Fig.\ 5 the generic result for  $\sigma_{\rm\scriptscriptstyle DC}(T)$ is
shown.
 In the homogeneous phase above $T_{\rm C}$ $\sigma_{\rm\scriptscriptstyle
DC}\approx
0.295$
is proportional to $1/T^2$. Just below  $T_{\rm C}$, the conductivity drops
because the DOS is reduced considerably due to the formation of the energy gap.
Unexpectedly, however, it does not vanish at $T=0$ but shows even a moderate
increase. This phenomenon results from a subtle cancellation
in the limit $T\to 0$ of an exponentially
 vanishing density of charge carriers (no spectral weight within the gap at
$T=0$) and a
diverging mobility due to the suppression of scattering processes
by the energy gap (infinite life time of quasi-particles)
\cite{mydiss,uhrig94c}.
In real physical systems, of course, this phenomenon will be obscured as soon
as other scattering processes than the interaction driven ones (e.g.\ those
driven by
disorder) become more important. Yet in very pure systems the part of
Fig.\ 5 at not too low temperatures should be observable.

Summarising, the feasibility of a high dimensional expansion even on the
two-particle level was shown for
interacting spinless fermions. Explicit results for the AC- and DC-conductivity
 in the homogeneous phase and in the AB-CDW were presented and discussed.

The author thanks Professors D.\ Vollhardt and E. M\"uller-Hartmann for helpful
discussions and comments. He acknowledges the financial support of the SFB 341
of the
DFG.

\newpage
\section*{Figure captions}
\begin{quote}
{\bf Fig.\ 1:}
Approximate functional $\Phi_{\rm N}$ exact to ${\cal O}(1/d)$. The
nearest-neighbour site
indices $i$ and $j$ are summed. The first diagram generates the Hartree-, the
second the Fock- and the third the local correlation terms.

{\bf Fig.\ 2:}
Order parameter $b(T)$ at $U=2$ in $d=3$; short dashed line: Hartree-, dashed
line:
Hartree-Fock-, solid line: complete $1/d$ result.

{\bf Fig.\ 3:}
Diagrammatic representation of the series for $\chi^{\sf JJ}$.

{\bf Fig.\ 4:}
 Real part of $\sigma_{\rm\scriptscriptstyle AC}$ in $d=3$ at $U=2$.
Curve a: $T=0.300$, $b=0.0$; curve b: $T=0.155$, $b=0.300$;
 curve c: $T=0.0$, $b=0.311$, $\delta$-peak at $\om= 1.133$ not displayed.

{\bf Fig.\ 5:}
 $\sigma_{\rm\scriptscriptstyle DC}(T)$ in $d=3$ at $U=2$.

\end{quote}


\begin{thebibliography}{10}

\bibitem{metzn89}
W.~Metzner, D.~Vollhardt, {\em Phys. Rev. Lett.} {\bf 62}, 324 (1989)

\bibitem{mulle89}
E.~M\"uller-Hartmann, {\em
  Z. Phys. B} {\bf 74}, 507 (1989)

\bibitem{vollh93}
D.~Vollhardt, in V.~J. Emery, ed., {\em Correlated Electron
  Systems}, p.~57, World Scientific, Singapore, (1993)

\bibitem{khura90}
A.~Khurana, {\em Phys. Rev. Lett.} {\bf 64}, 1990 (1990)

\bibitem{prusc}
T.~Pruschke, D.~L. Cox, M.~Jarrell, {\em Europhys. Lett.} {\bf 21}, 593 (1993);
 {\em Phys. Rev. B} {\bf 47}, 3553
  (1993)

\bibitem{halvo94}
E.~Halvorsen, G.~S. Uhrig, G.~Czycholl, {\em Z. Phys. B} {\bf 94}, 291 (1994)

\bibitem{mydiss}
G.~S. Uhrig, {\em Symmetriebrechung und Leitf\"ahigkeit f\"ur spinlose
  Fermionen in hohen Dimensionen}, vol.~11, series ABPKM, Verlag der Augustinus
Buchhandlung,
  Aachen, (1994)

\bibitem{baym}
G.~Baym, L.~P. Kadanoff, {\em Phys. Rev.} {\bf 124}, 287 (1961);
G.~Baym, {\em Phys. Rev.} {\bf 127}, 1391 (1962)

\bibitem{uhrig94c}
G.~S. Uhrig, to be published (1994)

\end{thebibliography}
\end{document}